\documentclass[10pt,conference]{ieeeconf}
\usepackage{graphics} 
\usepackage{epsfig} 
\usepackage{amsmath} 
\usepackage{amssymb} 
\usepackage{amsfonts}
\usepackage{graphicx}
\usepackage[normalem]{ulem}
\usepackage[labelfont=bf]{caption}
\usepackage[usenames,dvipsnames,table]{xcolor}
\usepackage{subcaption}
\usepackage{ulem}

\def\R{\mathbb{R}}
\def\>{\rightarrow}
\def\({\left(}
\def\){\right)}
\def\eps{\varepsilon}
\def\>{\rightarrow}
\newcommand{\mb}[1]{\mathbf{{#1}}}
\newcommand{\mc}[1]{\mathcal{{#1}}}

\IEEEoverridecommandlockouts

\begin{document}

\title{Patterned Dynamics of Delay-Coupled Swarms with Random Communication Graphs}
\author{K. Szwaykowska, L. Mier-y-Teran-Romero and I. B. Schwartz
\thanks{US Naval Research Laboratory, Code 6792, Nonlinear System Dynamics Section,
Plasma Physics Division, Washington, DC 20375. Email: klementyna.szwaykowska.ctr@nrl.navy.mil, lmieryt1@jhu.edu, ira.schwartz@nrl.navy.mil}}
\date{\today}

\maketitle


\begin{abstract}

Swarm and modular robotics are an emerging area in control of autonomous systems. However, 
coordinating a large
group of interacting autonomous agents requires careful consideration of the logistical issues
involved. In particular, inter-agent communication generally involves time delay, and bandwidth
restrictions limit the number of neighbors with which each agent in the swarm can communicate. In
this paper, we  analyze coherent pattern dynamics of groups of delay-coupled agents, where the
communication network is an Erd{\"o}s-Renyi graph. We show that overall motion patterns for a
globally-coupled swarm persist under decreasing network connectivity, and derive the bifurcation
structure scaling relations for the emergence of different swarming behaviors as a function of the
average network degree. We show excellent agreement between the theoretical scaling results and
numerical simulations.

\end{abstract}


The emergence of complex dynamical behaviors from simple rules of pairwise interaction in aggregates
of agents is a remarkable yet widespread phenomenon that appears in a multitude of application
domains. In biology, aggregates form at all spatial and temporal scales, from the microscopic (ex.,
aggregates of bacterial cells or the collective motion of skin cells in wound healing)
\cite{Budrene1995,Polezhaev2006,Lee2013} to large-scale systems of fish, birds, and even humans
\cite{Tunstrom2013,Helbing1995,Lee2006}. In control systems, the emerging capabilities of
swarms can allow groups of relatively small, inexpensive agents to achieve tasks that are beyond
any individual agent. For example, aggregates of locally interacting agents have been proposed as a
means to create scalable sensor arrays for surveillance and exploration
\cite{Bhatta2005,chuang2007,Lynch2008,Yang2008,Wu2011,Wu2012};
distributed sensing \cite{Chung2006,Kar2008}; cooperative construction
\cite{Augugliaro2014,Werfel2014}; and the formation of reconfigurable modal systems
\cite{Bandyopadhyay2014,Dorigo2013,Rubenstein2014}.

The application of multi-agent systems for various tasks occurs in parallel with a modeling effort
aimed at understanding the emergent properties of swarms. Existing literature on the subject 
provides a great selection of both individual-based 
\cite{Helbing1995,Lee2006,Vicsek2006,Tunstrom2013} and continuum models
\cite{Edelstein-Keshet1998,Topaz2004,Polezhaev2006} of group motion. Many biologically-inspired 
models are based on ``zones'' of attraction, repulsion, and/or alignment interactions between 
agents \cite{Miller2012,Tarras2013,Viragh2014}. Setting different ranges for attraction/repulsion 
interactions between all agents in the swarm or introducing heterogeneous dynamics can lead to 
interesting behaviors, such as splitting of a large swarm into smaller groups 
\cite{Kumar2010,Chen2010,Chen2011}. Swarming with nearest-neighbor alignment is studied 
numerically in \cite{Vicsek1995}. Stochastic interactions between agents are modeled in 
\cite{Nilsen2013}.

Even so, most swarming models do not lend themselves to analysis, and there
is still limited understanding of how group behaviors and structures arise out of simple 
inter-agent interactions. 
However, some progress has been made in understanding how the structure and parameters of a given 
swarm model contribute to the aggregate motion of the group. For example, \cite{Viscido2005} 
presents a simulation-based analysis of the different kinds of motion in a fish-schooling model; 
the authors map phase transitions between different aggregate behaviors as a function of group size 
and maximum number of neighbors that influence the motion of each ``fish''. In \cite{Lee2006}, the 
authors use simulations to study transitions in aggregate motions of prey in response to a predator 
attack.

In this paper, we are motivated by the idea of using a swarm of interacting autonomous aerial robots
to conduct surveillance/monitoring of a specified environment, where a human operator uses a 
high-level control to guide the swarm as a whole (e.g., to set the monitoring region), while 
individual agent trajectories are governed by swarm interactions. The idea is similar to the 
reduced-order swarm control described in \cite{Belta2004,Michael2006} and references therein. Our 
goal is to rigorously characterize the swarm motion patterns under a simple but general swarming 
model, as a function of model parameters, so as to exploit them for parametric control of the 
system. We incorporate two key model modifications for real-world applicability: time delay and 
restricted communication bandwidth. 

Systems of interacting individual agents, whether natural and engineered, involve some degree of 
communication delay \cite{Martin2001,Bernard2004,Monk2003,Forgoston2008}. Time delay can have 
significant impact on system dynamics, leading to destabilization or synchronization of coupled 
systems \cite{Papachristodoulou2006,Zuo2010,Englert2011}. As shown in our earlier work with 
globally delay-coupled swarms of homogeneous \cite{Romero2011,Romero2012,Lindley2013a} and 
heterogeneous agents \cite{Szwaykowska2014}, communication delay can cause emergence of new 
collective motion patterns and, in the presence of noise, lead to switching between bistable 
patterns; this, in turn, can lead to instability in robotic swarming systems 
\cite{Viragh2014,Liu2003}. 

Because global coupling is easier to analyze and a reasonable assumption in cases of high-bandwidth 
communication and when the number of agents is small, many models make the mathematically 
simple but physically implausible assumption that swarms are globally coupled (that is, each agent 
is influenced by the motion of all other agents in the swarm) 
\cite{Motsch2011,Chen2014a,Lee2006,Vecil2013a}. However, for large groups of agents, bandwidth 
restrictions generally mean that it is not feasible to maintain all-to-all communication. In this 
paper, we generalize our previous results for globally-coupled swarms by assuming an 
Erd{\"o}s-Renyi communication network, where the mean number of neighbors for each agent in 
the swarm is fixed. 

We use mean-field dynamics to analytically predict transitions between regimes of different 
collective swarm motions as a function of model parameters for swarms consisting of homogeneous 
delay-coupled agents with a fixed undirected Erd{\"o}s-Renyi communication network. We use a simple yet 
general particle swarming model, which combines agent self-propulsion and inter-agent attraction. 
This choice can be justified in the case of swarming robots with very fast relaxation times, such 
as quadcopters, which can be treated as holonomic vehicles over the spatio-temporal scales required
for a surveillance/monitoring deployment. We show that the stable motion 
patterns observed in the globally-coupled system (translation, ring formation, and rotation about a 
common center of mass) \cite{Romero2012} persist under non-global coupling, for appropriately 
chosen model parameters. Our results are verified through numerical simulations.

\section{Problem Statement}

Consider a swarm of delay-coupled agents in the plane, with each agent indexed by $i$. We build an Erd{\"o}s-Renyi 
communication network on the swarm as follows. Starting with a globally-coupled network, pick an 
existing link at random, using a uniform distribution over the existing links, and remove it, until 
the mean degree matches some target value. Let $\mc{N}_i$ denote the set of neighbors 
of agent $i$, that is, the set of agents with which agent $i$ shares a communication link. Because 
of communication/sensing delays, information about the position of the neighboring agents is 
available with delay $\tau$, assumed to be the same for all agents.  For simplicity, we assume that all agents in the swarm have identical 
dynamics and that all connections in the network are bidirectional and fixed in time. The agents 
have self-propulsion and are attracted to their neighbors with strength that depends on the coupling 
coefficient $a$. Then the motion of agent $i$ is governed by the following equation:
\begin{equation}
 \ddot{\bold r}_i = (1-|\dot{\bold r}_i|^2)\dot{\bold r}_i - \frac{a}{N}\sum_{j \in \mc{N}_i}({\bold r}_i(t) - {\bold r}_j^\tau(t)),
\end{equation}\label{Eq:agenti}
where superscript $\tau$ is used to denote time delay, so that 
${\bold r}_j^\tau(t) = {\bold r}_j(t-\tau)$, and $|\cdot|$ denotes the Euclidean norm. The
first term in the above equations represents the self-propulsion of swarm agents. The second term models
pairwise attraction between each agent and its neighbors in the swarm. This simplified model does
not include short-range repulsion or other collision-avoidance strategies; however, earlier studies
indicate that the collective dynamics of the swarm are not significantly altered by the introduction
of short-range repulsion terms \cite{Romero2012}.

We examine the dynamics of the system analytically in the limit where the system is
almost completely connected ($\frac{(N-1)-card(\mc{N}_i)}{N-1} \ll 1$, where
$card(\mc{N}_i)$ is the number of neighbors of node $i$), and show via simulations that the approximations made in the almost-connected limit hold closely even as the mean coupling
degree is reduced to less than $50\%$ of possible links. 

\section{System Dynamics in the Mean-Field}

In \cite{Romero2012} we derived the mean-field dynamics of the system in the limit 
$N \> \infty$ where the delayed coupling was considered to be all-to-all. The difference considered 
in the model given by Eq.~\ref{Eq:agenti} is that each agent's motion depends only on its 
neighbors, rather than all other agents in the swarm. Let 
${\bold R}(t) = \frac{1}{N} \sum_{i=1}^N {\bold r}_i(t)$ denote the position of the center of mass 
of the swarm, and let $\delta {\bold r}_i(t) = {\bold r}_i(t) - {\bold R}(t)$;
note that $\sum_{i=1}^N \delta{\bold r}_i=0$. Let 
$\langle \cdot, \cdot \rangle$ denote the dot product in $\R^2$. The motion of the center of mass 
is governed by
\begin{equation}
\begin{split}
\ddot{\bold R}
&= (1-|\dot{\bold R}|^2) \dot{\bold R} \\
&\quad - \frac{1}{N}\sum_{i=1}^N \left( |\dot{\bold R}|^2 \delta \dot{\bold r}_i + (|\delta \dot{\bold r}_i|^2 + 2
\langle \dot{\bold R}, \delta \dot{\bold r}_i \rangle) (\dot{\bold R} + \delta \dot{\bold r}_i) \right) \\
&\quad - \frac{a \bar{p}}{N}({\bold R} - {\bold R}^\tau) - \frac{a}{N^2} \sum_{i=1}^N \left( p_i \delta {\bold r}_i -
\sum_{j \in \mc{N}_i} \delta {\bold r}_j^\tau \right),
\end{split}
\end{equation}
where $p_i = card(\mc{N}_i)$ is the number of neighbors of agent $i$ (in the globally coupled case, $p_i = N-1$) and $\bar{p} = \frac{1}{N} \sum_{i=1}^N p_i$ is the mean degree of the network. We denote the 
mean fraction of missing links by $\eps$, so that $\bar{p} =
(1-\eps)(N-1)$. Note that, if $p_i = p$ is 
the same for all $i$,
\begin{equation}
\sum_{i=1}^N \!\!\left(\! p \delta {\bold r}_i -\sum_{j \in \mc{N}_i} \delta
  {\bold r}_j^\tau \!\right) = - \sum_{i=1}^N \sum_{j \in \mc{N}_i} \delta
\mb{r}_j^\tau = - p \sum_{i=1}^N  \delta {\bold r}_i^\tau=0
\end{equation}
since $\sum_{i=1}^N \delta \mb{r}_i = 0$ and in the double summation, each
$\delta \mb{r}_i$ appears $p$ times because the
network is undirected and each node has $p$ neighbors. However, in our network
$p_i$ is drawn from a binomial distribution. Our previous work shows that in
many instances either individual deviations from the center of mass are
small, or in aggregate they average out over the whole population. (We
discuss situations in which this assumption breaks in a later section). Then, neglecting all terms of order $\delta
\mathbf{r}_i$ and in the limit $N \> \infty$, the center of mass motion is given approximately by
\begin{equation} \label{eq:ddotR}
 \ddot{\bold R} = (1-|\dot{\bold R}|)\dot{\bold R} - a (1-\varepsilon)({\bold R}(t) - {\bold R}^\tau(t)).
\end{equation}

We note that the mean-field dynamics are identical to those for the globally-coupled homogeneous
swarm described in \cite{Romero2012}, with coupling constant $a$ replaced by an `effective' value 
$a_\text{eff} = a(1-\eps)$. Quite remarkably, simulation results indicate that 
the system exhibits similar collective motions to the globally coupled case, even as $\bar{p}$ is
significantly decreased. These basic collective motion patterns are ``translation'', where the entire swarm
collapses to a point and travels along a straight-line trajectory at constant speed; ``ring''
motion, where the swarm agents form concentric counter-rotating rings about the stationary center of
mass; and ``rotating'' motion, where the agents collapse to a small volume and collectively rotate about a
fixed point. The collective motions of the swarm and the effects of non-global coupling are
described in more detail in the following section.

\section{Collective swarm motions}

As in the globally coupled case, the steady-state motions of the swarm depend on values of the
coupling coefficient $a$, the delay $\tau$; in addition, there is now a dependence on the fraction
of missing links $\eps$. The collective motion patterns of the swarm for different values of the 
parameters $a$ and $\tau$ are described in more detail below.

\subsection{Translating state}
In the translating state, the agent locations all lie close to the swarm center of mass, and the
swarm moves with constant speed and direction. 
Following the calculation in \cite{Romero2012}, it can be shown that the translation speed $\|\dot{\mb{R}}\|$ must satisfy $\|\dot{\mb{R}}\|^2 = 1-a_\text{eff}\tau$. The
system exhibits a pitchfork bifurcation at $a_\text{eff}\tau = 1$, where the translating state disappears.

\subsection{Ring state}
\begin{figure}[htb]
\centering
\begin{subfigure}[b]{0.4\textwidth}
 \includegraphics[width=\textwidth]{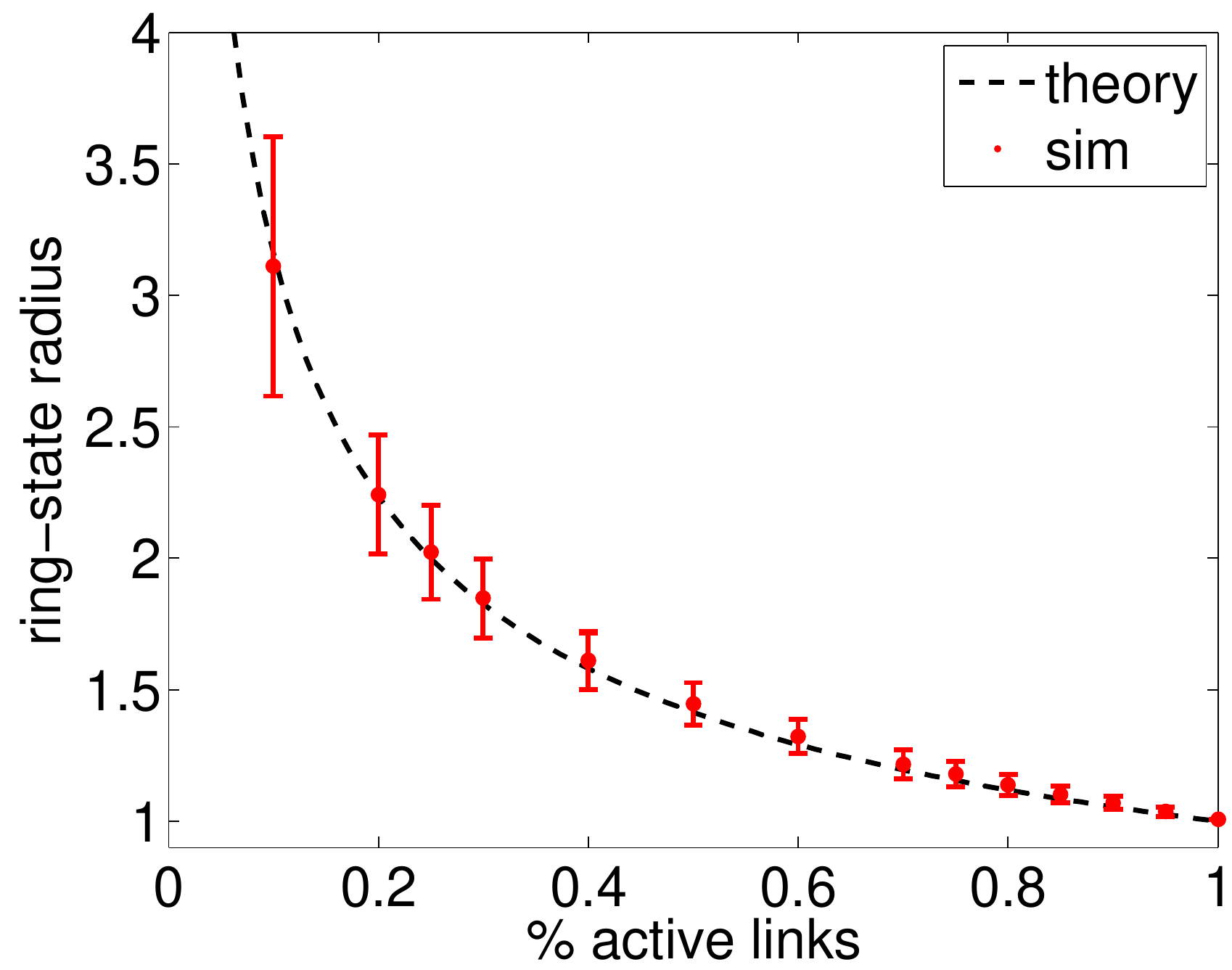}
\end{subfigure}
\begin{subfigure}[b]{0.4\textwidth}
 \includegraphics[width=\textwidth]{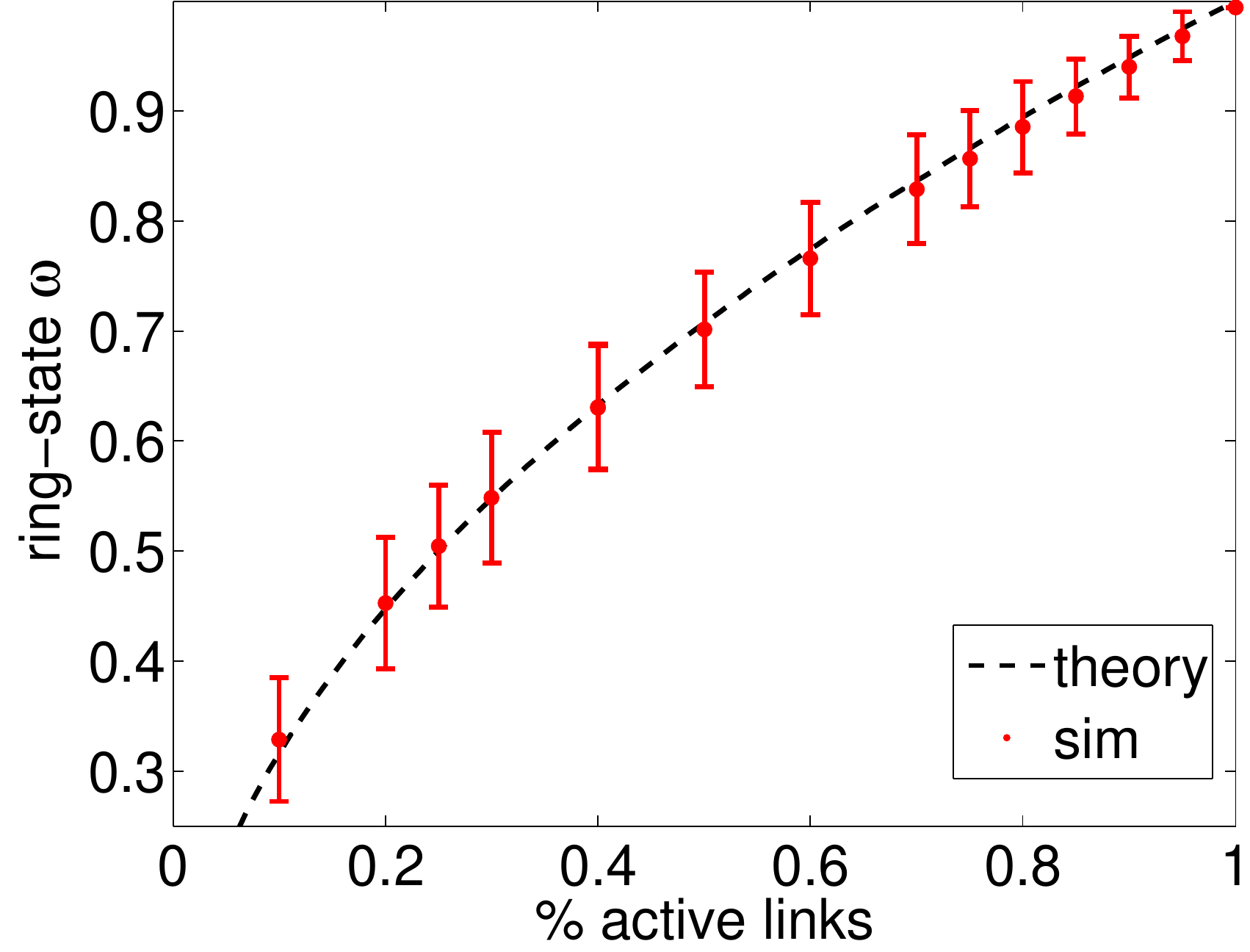}
\end{subfigure}
\caption{Comparison of the theoretical values of radius (top) and angular velocity (bottom) of the
agents in the ring state, for different connection degrees, compared with full-swarm
simulation of 150 agents. The x-axis represents the percentage of active links, out of all possible
links; all links are bidirectional and time-invariant. Error bars are shown one standard deviation
above and below the mean values. The simulations were run with $a = 1$ and $\tau = 2.5$.}
\label{fig:ringstate}
\end{figure}

For all values of $a$ and $\tau$, (\ref{eq:ddotR}) admits a stationary
solution, $\mb{R}(t) = \mb{R}(0)$. In this state, the agents converge to a pair of concentric, counter-rotating rings about the stationary center of mass. The mean radius and angular velocity of the agents in the ring state satisfy
\begin{align}
 \rho &= 1/\sqrt{a_\text{eff}} \\
 \omega &= \pm \sqrt{a_\text{eff}},
\end{align}
respectively (see Fig.~\ref{fig:ringstate}). The stability of the ring state
is determined by the characteristic equation
\begin{equation}
 M(\lambda; a,\tau) = \Big[(\lambda^2 - \lambda + a_\text{eff})^2 - (\lambda^2 - \lambda + a_\text{eff})a_\text{eff} e^{-\lambda \tau}\Big]^2
\end{equation}
and is lost along Hopf bifurcation curves which give rise to a rotating state.

\subsection{Rotating state}

\begin{figure}[htb]
 \centering
 \includegraphics[width=.4\textwidth]{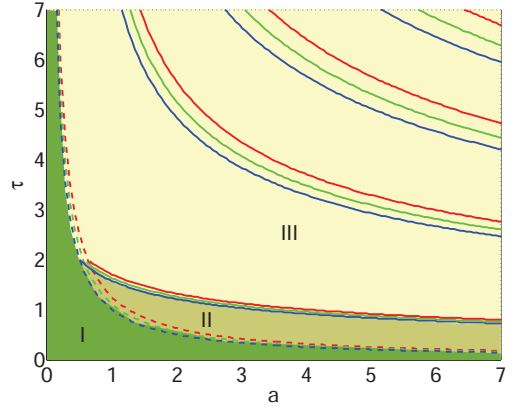}
 \caption{Hopf bifurcation curves for $\eps=0$ (blue), $\eps=0.1$ (green), and $\eps=0.2$ (red).
The dashed lines show the location of the pitchfork bifurcation where the translating state
disappears. In the mean field, the translating state occurs in region I; the ring state in region II, and the 
rotating state, in region III.}
 \label{fig:avtau}
\end{figure}

Solving for values where roots of $M$ cross the imaginary axis, we obtain Hopf bifurcation curves in the $a$ and $\tau$ plane:
\begin{equation}
 \tau = \frac{1}{\sqrt{2a_\text{eff}-1}}\left( \arctan \frac{\sqrt{2a_\text{eff}-1}}{1-a_\text{eff}} +
 2 m \pi\right),
\end{equation}
$m \in \mathbb{Z}$. When $\eps=0$, we recover the equations for the globally
coupled system by taking $a_\text{eff}=a$; the
factor of $(1-\eps)$ in $a_\text{eff}$ which results from breaking a fraction of the links in the global network
represents a perturbation from the globally-coupled case (Fig.~\ref{fig:avtau}). Note that the pitchfork and 
Hopf bifurcation curves meet at a Bogdanov-Takens point when $a = \frac{1}{2(1-\eps)}$, $\tau = 2(1-\eps)$. 

\begin{figure}[htb]
\centering
\includegraphics[width=0.4\textwidth]{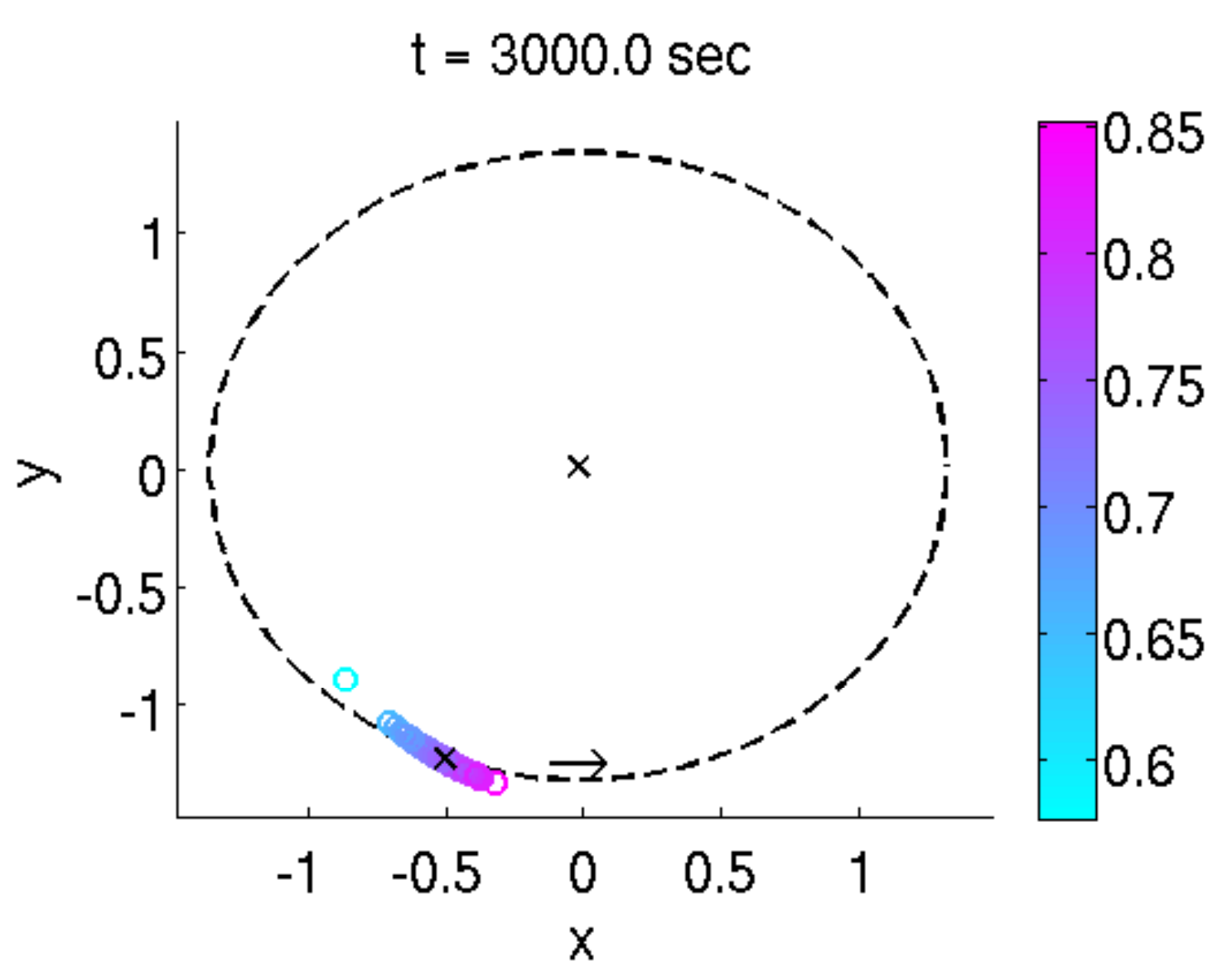}
\caption{Snapshot of simulation showing swarm in rotating state about a stationary center point
(marked by `$\times$'). The colors indicate the coupling degree of each agent, normalized by $N-1$.
Here $a = 1$, $\tau = 4.5$, and $\bar{p}/(N-1) = 0.75$.}
\label{fig:rotex}
\end{figure}

\begin{figure}[htb]
\centering
\begin{subfigure}[b]{0.4\textwidth}
 \includegraphics[width=\textwidth]{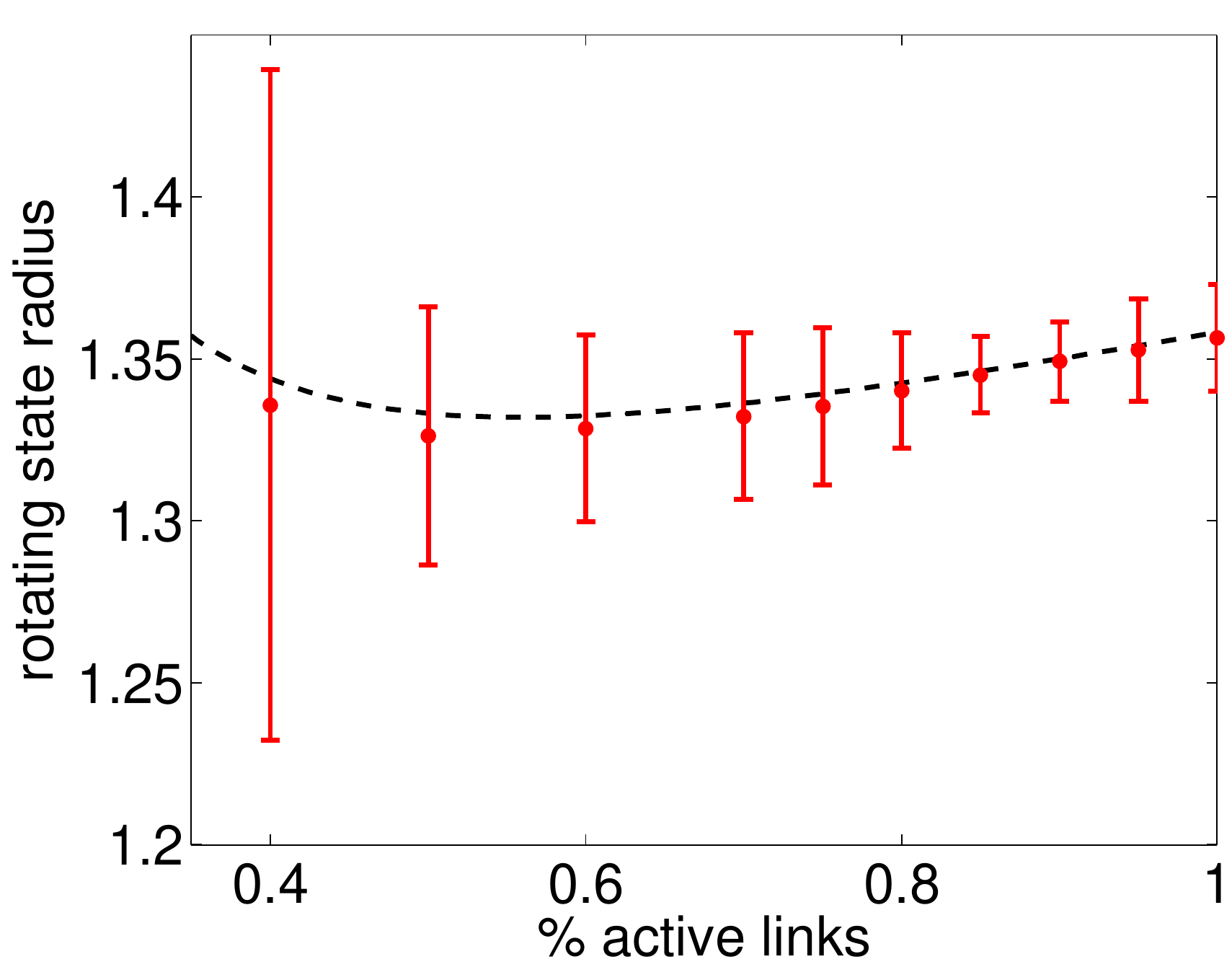}
\end{subfigure}
\begin{subfigure}[b]{0.4\textwidth}
 \includegraphics[width=\textwidth]{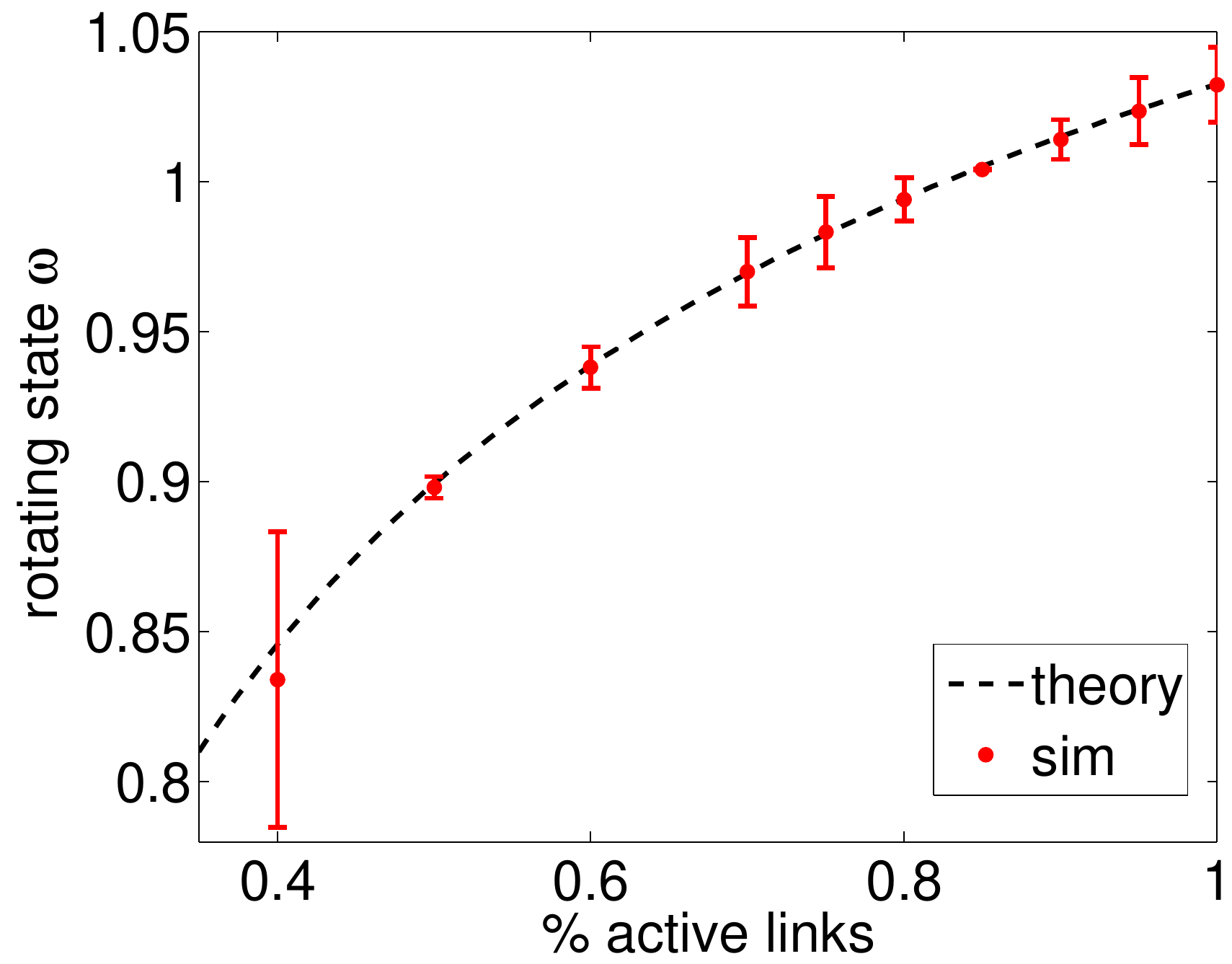}
\end{subfigure}
\caption{Comparison of the theoretical values of radius (top) and angular velocity (bottom) of the
agents in the rotating state with full-swarm simulations of 150 agents, for different connection 
degrees. The simulation values are obtained by averaging over all agents int he swarm. The x-axis 
represents the percentage of active links, out of all possible links; all links are bidirectional. 
Error bars are shown one standard deviation above and below the mean.}
\label{fig:rotstate}
\end{figure}

The rotating state, in which the agents collapse and collectively rotate about a fixed point is created at the first Hopf bifurcation.
In the case of global
coupling, all agent positions coincide; however, when coupling is not global, different agents
circle the fixed point with equal angular frequency but have different radii, and have a fixed relative
phase offset from the center of mass, depending on their coupling degree (see Fig.
\ref{fig:rotex}). The radius and angular velocity of the center of mass of
the swarm in the rotating state satisfy
\begin{align}
 \rho  &= \frac{1}{|\omega|}\sqrt{1-a_\text{eff} \frac{\sin \omega\tau}{\omega}} \\
 \omega^2 &= a_\text{eff}(1-\cos \omega \tau)
\end{align}
(see Fig.~\ref{fig:rotstate}).

We now investigate the precise spatial organization of swarming agents in the rotating state. Using $\mathbf{r}_j^\tau= \mathbf{R}^\tau + \delta \mathbf{r}_j^\tau$ allows us to write the coupling of agent $i$ to the other particles as
\begin{align}
\sum_{j \in \mc{N}_i}({\bold r}_i(t) - {\bold r}_j^\tau(t)) =& \ p_i (\mathbf{r}_i - \mathbf{R}^\tau) - \sum_{j \in \mc{N}_i}\delta \mathbf{r}_j^\tau \notag\\
\approx & \ p_i (\mathbf{r}_i - \mathbf{R}^\tau),
\end{align}
after neglecting the sum of small terms $\delta \mathbf{r}_j^\tau$. In this way, the dynamics of agent $i$ becomes
\begin{equation}
 \ddot{\bold r}_i = (1-|\dot{\bold r}_i|^2)\dot{\bold r}_i - \frac{a p_i}{N}({\bold r}_i(t) - {\bold R}^\tau(t)),
\end{equation}
so that dynamically, less-than-global coupling in our Erd{\"o}s-Renyi network
is approximately equivalent to a fully connected network of agents with
heterogeneous coupling coefficients
\begin{equation}
a_i = a p_i/(N-1),
\end{equation}
where the factor of $N-1$ is introduced so that $a_i = a$ in the fully
connected case where $p_i = N-1$. In the limit $N \> \infty$, we have
\begin{equation}
 \ddot{\bold r}_i = (1-|\dot{\bold r}_i|^2)\dot{\bold r}_i - a_i \left( {\bold r}_i - {\bold R}^\tau \right), \label{eq:ddotrglobal}
\end{equation}
where the motion of the center of mass $\mb{R}$ is given by (\ref{eq:ddotR}).

Let $(\rho, \theta)$ and $(\rho_i,\theta_i)$ denote the polar coordinates of the swarm center of mass 
and of agent $i$, respectively. In these coordinates, the equations of motion for the swarm are
\begin{align}
\ddot{\rho} &= (1 - \rho^2\dot{\theta}^2 - \dot{\rho}^2)\dot{\rho} + \rho \dot{\theta}^2 -
\bar{a}(\rho - \rho^\tau \cos(\theta-\theta^\tau)) \\
\rho \ddot{\theta} &= (1 - \rho^2\dot{\theta}^2 - \dot{\rho}^2) \rho \dot{\theta} -
2\dot{\rho}\dot{\theta} - \bar{a}\rho^\tau\sin(\theta-\theta^\tau) \\
\ddot{\rho}_i &= (1 - \rho_i^2\dot{\theta}_i^2 - \dot{\rho}_i^2)\dot{\rho}_i + \rho_i
\dot{\theta}_i^2 - a_i(\rho_i - \rho^\tau \cos(\theta_i-\theta^\tau)) \\
\rho_i \ddot{\theta}_i &= (1 - \rho_i^2\dot{\theta}_i^2 - \dot{\rho}_i^2) \rho_i \dot{\theta}_i -
2\dot{\rho}_i\dot{\theta}_i - a_i\rho^\tau\sin(\theta_i-\theta^\tau),
\end{align}
where $\bar{a} = \frac{1}{N}\sum_{i=1}^N a_i$ is the mean coupling
coefficient.

\begin{figure}[htb]
\centering
\begin{subfigure}[b]{0.4\textwidth}
  \includegraphics[width=\textwidth]{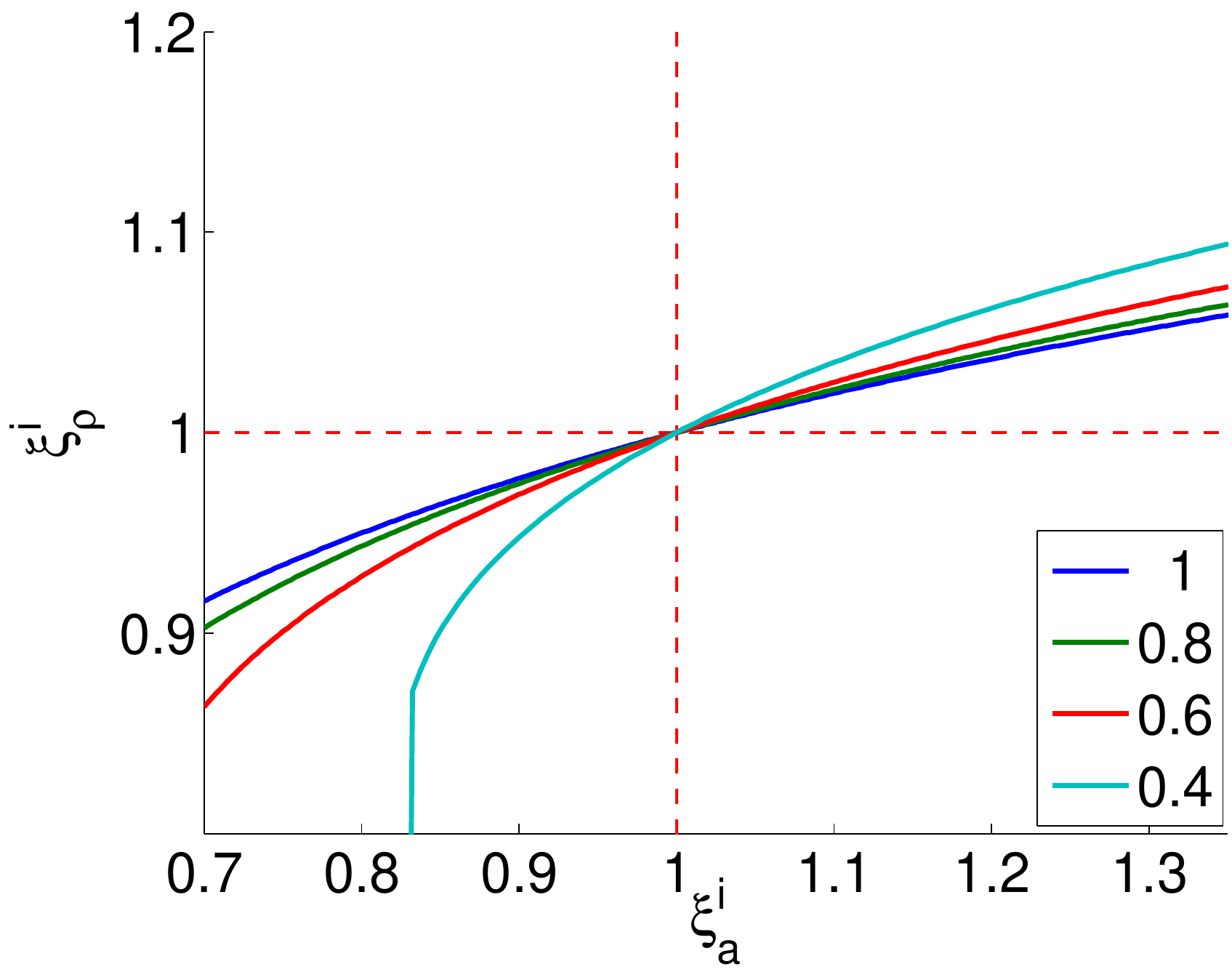}
\end{subfigure}
\begin{subfigure}[b]{0.4\textwidth}
  \includegraphics[width=\textwidth]{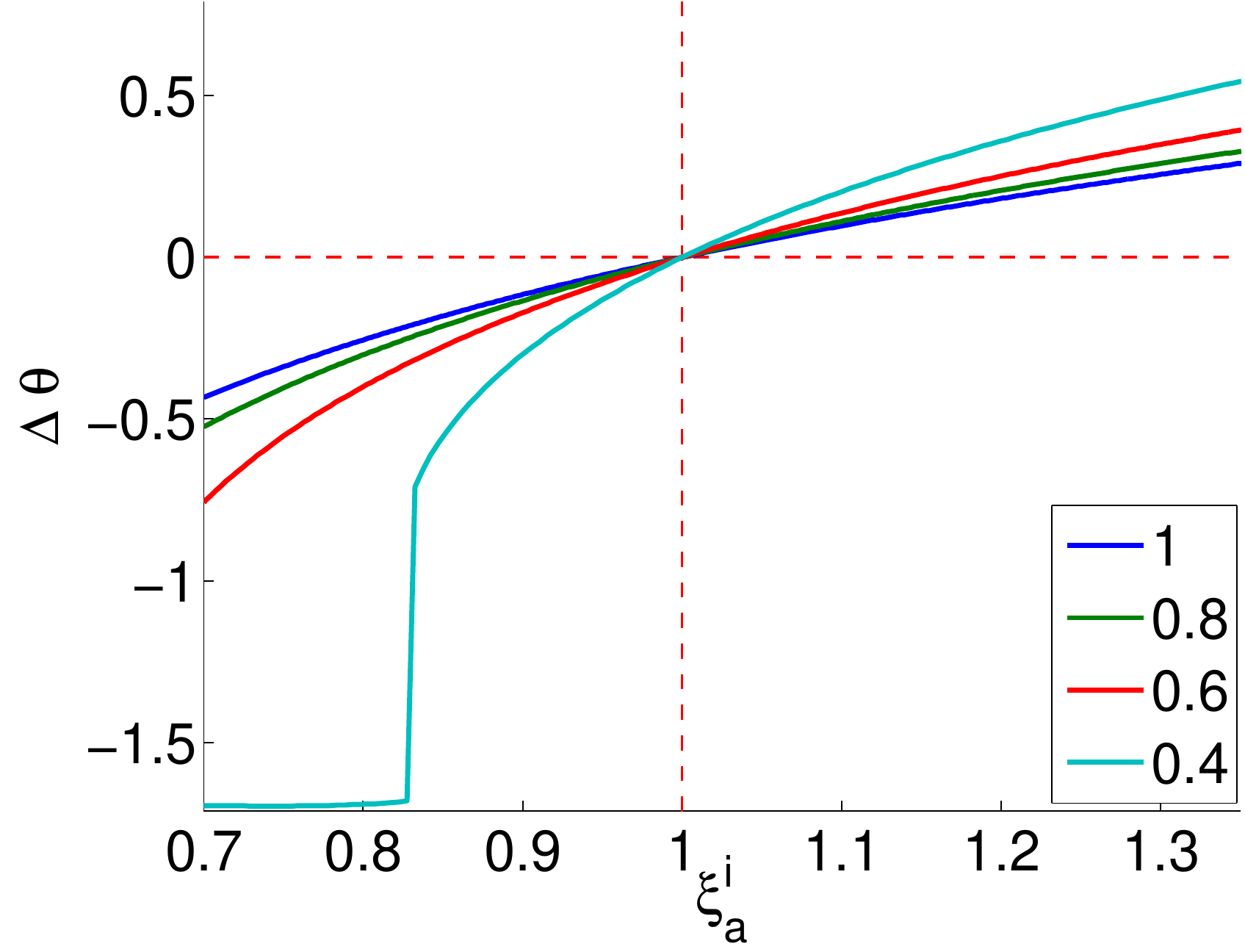}
\end{subfigure}
\caption{Theoretical values for ratio of radius for agent $i$ to radius of swarm center of
mass (top) and phase difference (bottom) as a function of $\xi_a^i = a_i/\bar{a}$, for $\bar{a} = 1,\,0.8,\,0.6$, and $0.4$.}
\label{fig:twopopthm}
\end{figure}

\begin{figure}[htb]
\centering
\begin{subfigure}[b]{0.23\textwidth}
\includegraphics[width=\textwidth]{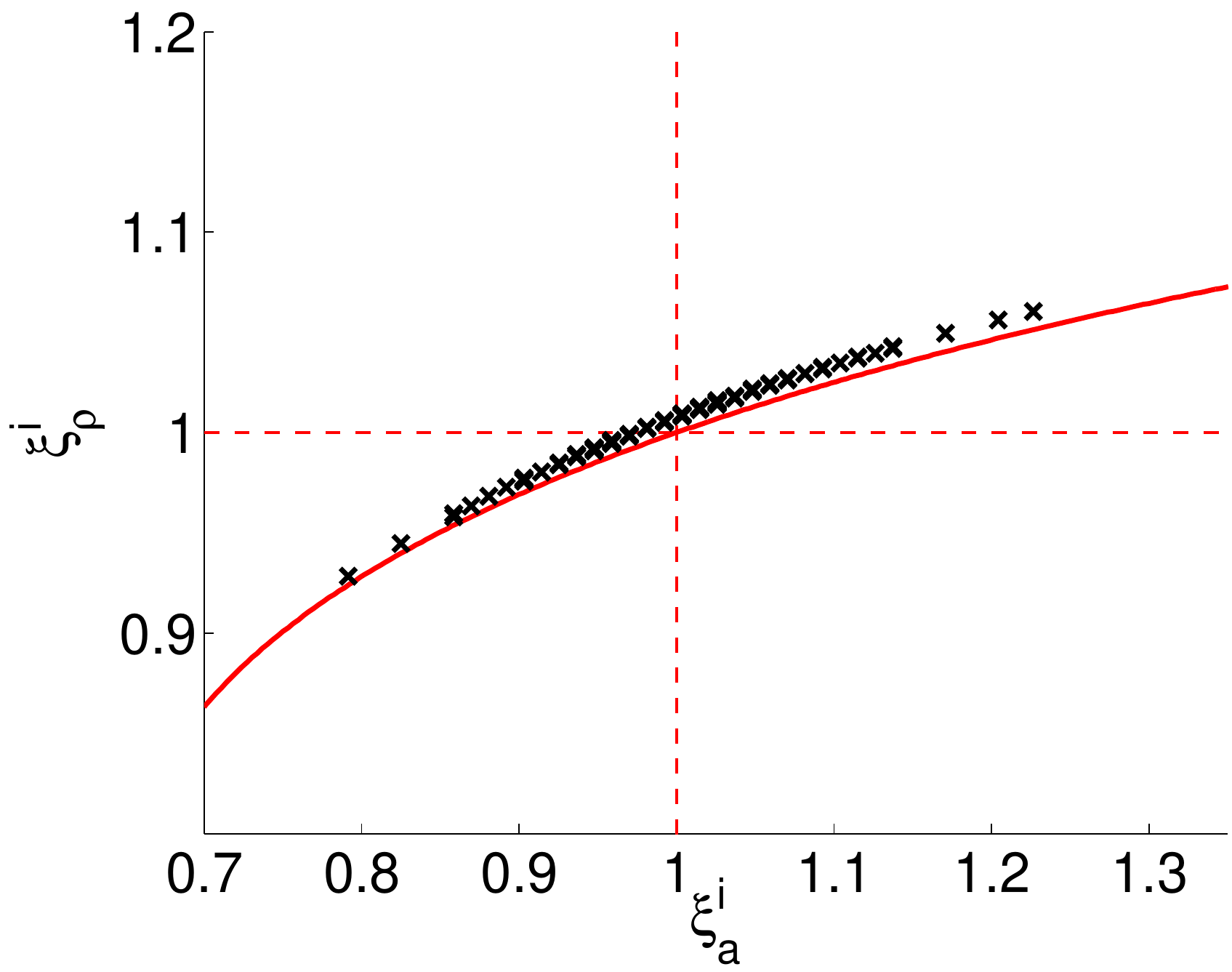}
\end{subfigure}
\begin{subfigure}[b]{0.22\textwidth}
\includegraphics[width=\textwidth]{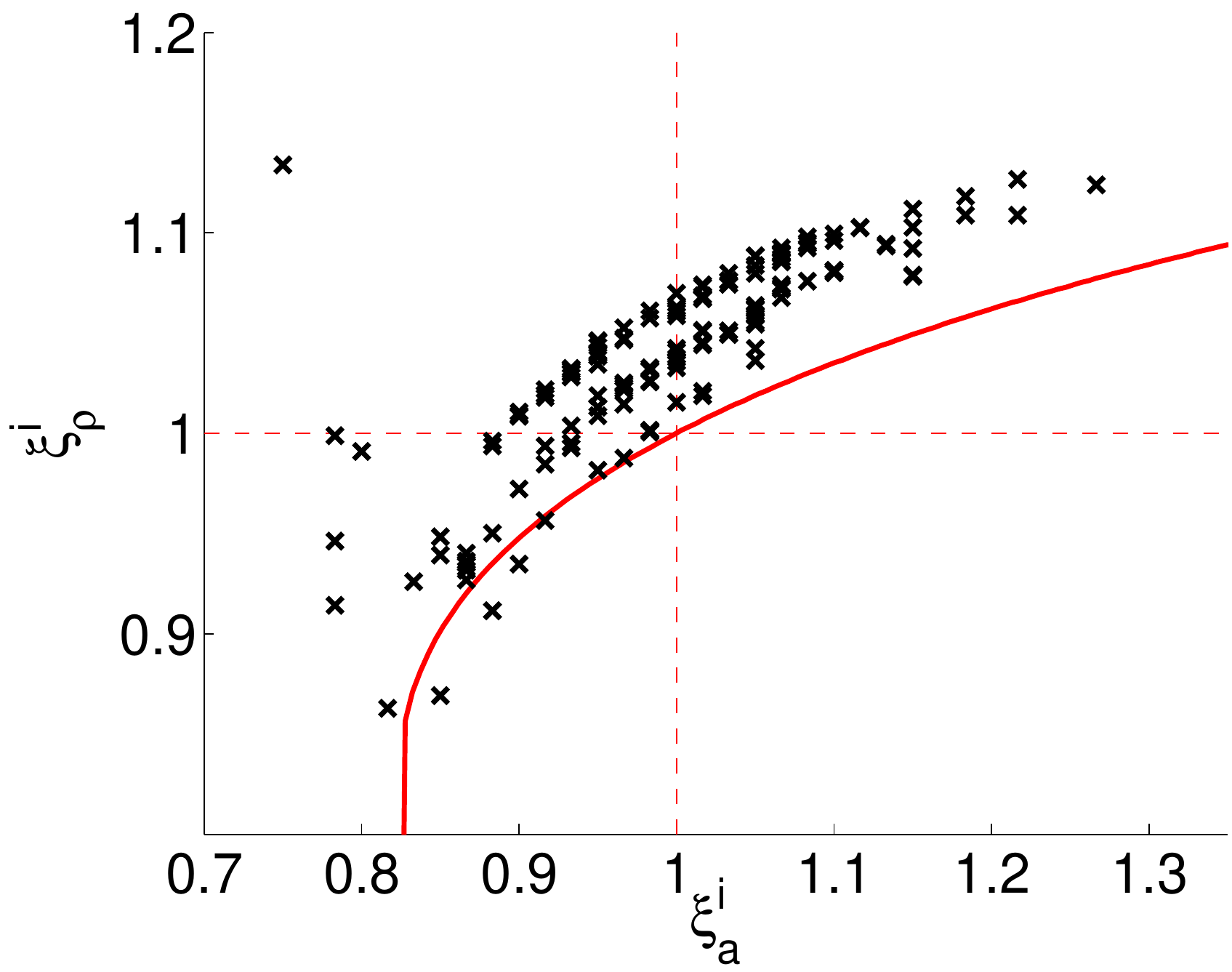}
\end{subfigure}
\begin{subfigure}[b]{0.23\textwidth}
\includegraphics[width=\textwidth]{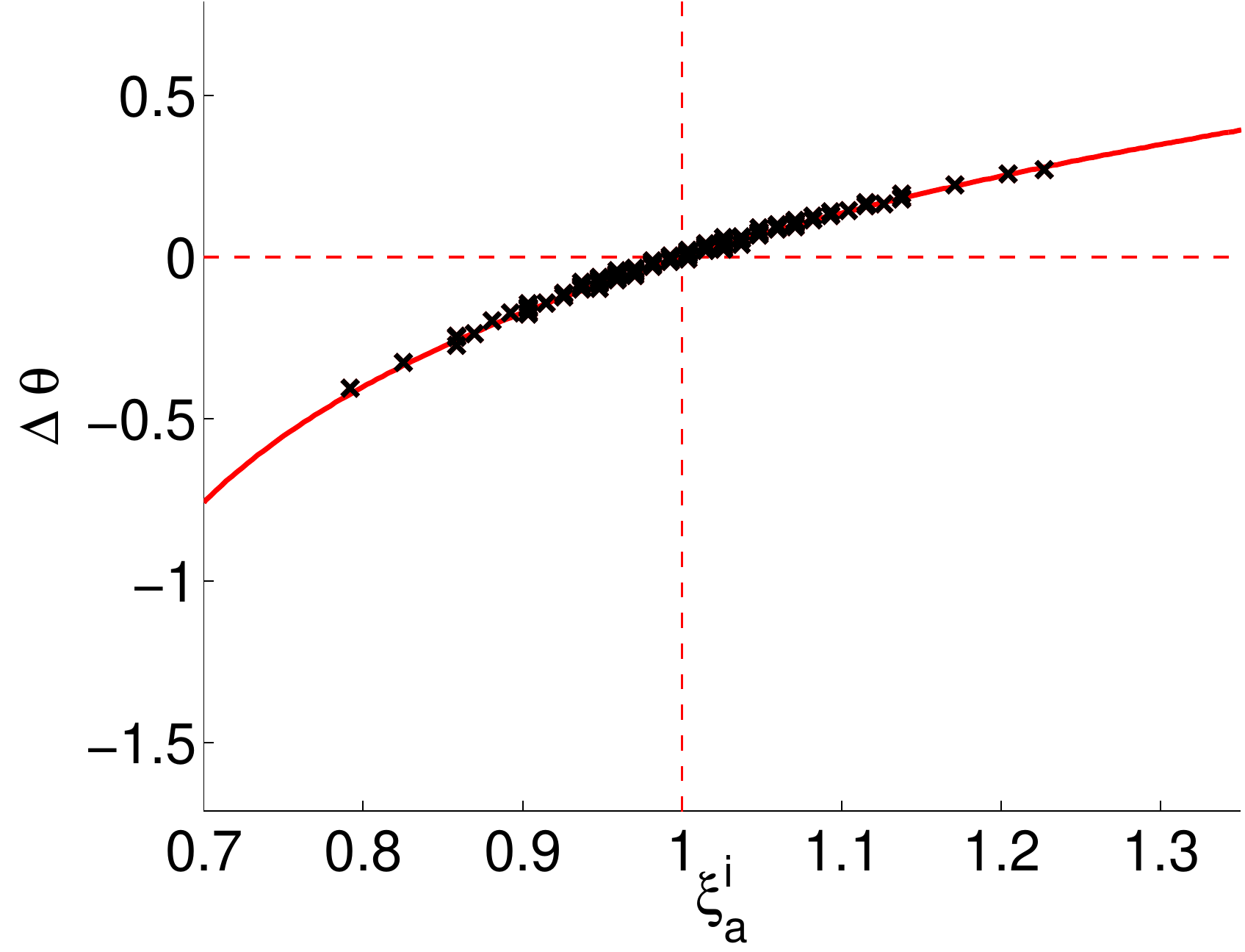}
\end{subfigure}
\begin{subfigure}[b]{0.23\textwidth}
\includegraphics[width=\textwidth]{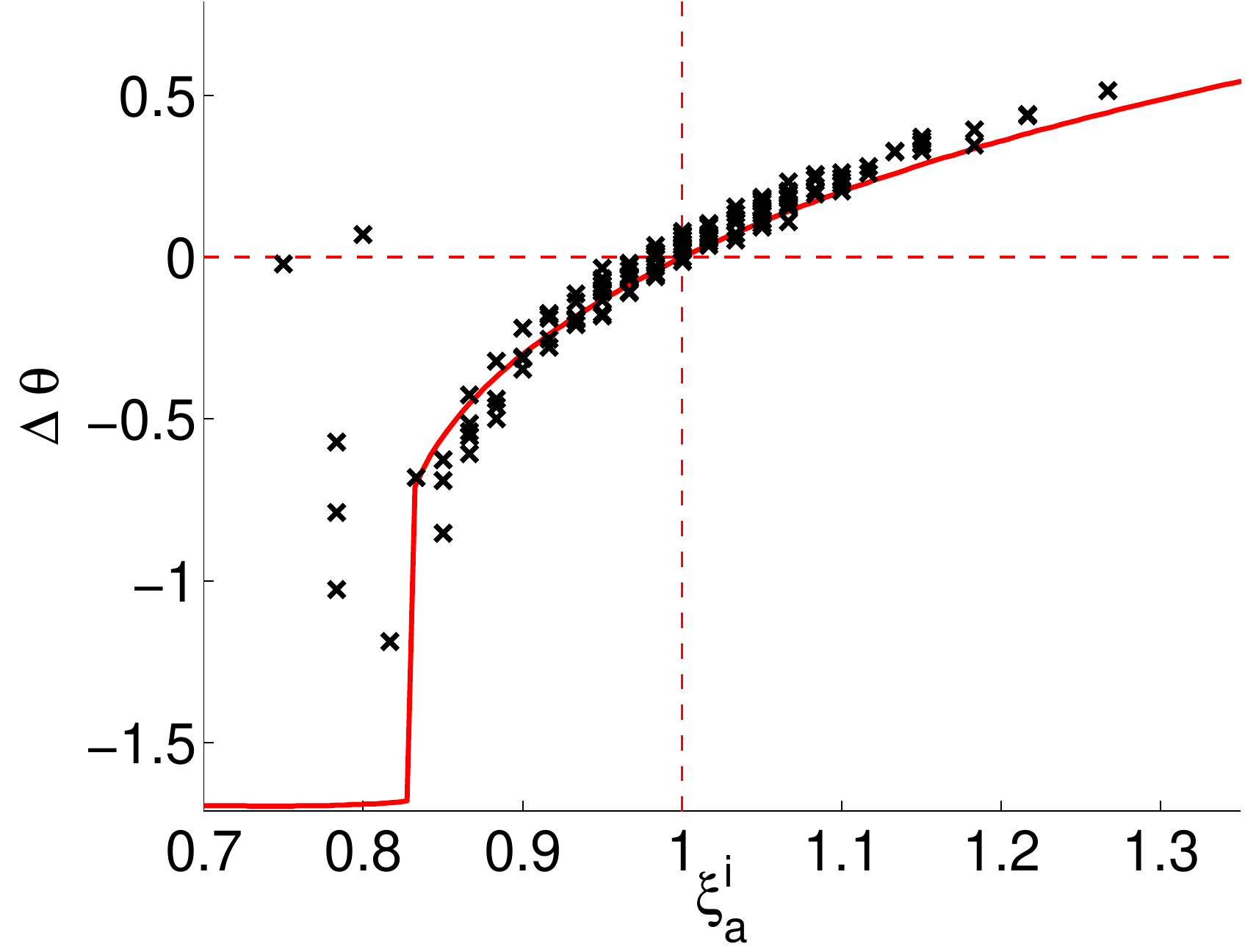}
\end{subfigure}
\caption{Comparison of simulation and theoretical values for radius (top) and phase difference 
from the center of mass (bottom) for agent $i$. Results are shown for two swarms, the first with 
$\bar{a} = 0.6$ (left) and the second with $\bar{a} = 0.4$ (right) (that is, $60\%$ of links active 
and $40\%$ of links active, respectively). For both swarms, $\tau = 4.5$ sec, and number of agents 
is $150$. All links are bidirectional. Note that within each swarm, agents with higher coupling 
degree lie further from the center of rotation.}
\label{fig:twopopthm2}
\end{figure}

Without loss of generality, we set the origin of the polar coordinate system at the center of
rotation so that $\rho$ and $\rho_i$ are constant in time. Also, we have $\ddot{\theta} = \ddot{\theta}_i \equiv 0$. Let
$\omega = \dot{\theta}$ and $\omega_i = \dot{\theta}_i$. For convenience, we define
$\xi_a^i = a_i/\bar{a}$ and $\xi_\rho^i = \rho_i/\rho$. The equations of motion then become
\begin{align}
0 &= \rho \omega^2 - \bar{a}\rho(1 - \cos(\omega \tau)) \\
0 &= (1 - \rho^2\omega^2) \rho \omega - \bar{a}\rho\sin(\omega \tau) \\
0 &= \xi_{\rho}^i\rho \omega_i^2 - \xi_a^i \bar{a} \big(\xi_{\rho}^i \rho \notag\\
&\qquad - \rho \cos((\omega_i-\omega)t+\theta_i(0)-\theta(0) + \omega \tau)\big)
\label{eq:ddotrhoi} \\
0 &= (1 - (\xi_\rho^i)^2\rho^2\omega_i^2) \xi_{\rho}^i\rho \omega_i \notag\\
&\qquad - \xi_a^i \bar{a}\rho\sin((\omega_i-\omega)t+\theta_i(0)-\theta(0) + \omega \tau)).
\label{eq:ddotthetai}
\end{align}
Note that equations (\ref{eq:ddotrhoi}) and (\ref{eq:ddotthetai}) can only be satisfied for all
times $t$ if $\omega_i = \omega$. Let $\Delta \theta_i = \theta_i(0)-\theta(0)$ denote the angular offset between particle $i$ and the center of mass. Simplifying, we finally have
\begin{align}
0 &= \omega^2 - \bar{a}(1 - \cos(\omega \tau)) \label{eq:wsqr}\\
0 &= (1 - \rho^2\omega^2) \omega - \bar{a} \sin(\omega \tau) \label{eq:wrho}\\
0 &= \xi_{\rho}^i \omega^2 - \xi_a^i \bar{a}(\xi_{\rho}^i - \cos(\omega \tau + \Delta \theta_i)) \\
0 &= (1 - (\xi_\rho^i)^2\rho^2\omega^2) \xi_{\rho}^i \omega - \xi_a^i \bar{a} \sin(\omega \tau +
\Delta \theta_i)).
\end{align}
This set of coupled nonlinear equations can be solved numerically for $\rho$, $\omega$,
$\xi_{\rho}^i$, and $\Delta \theta_i$ for different values of $\bar{a}$, $\tau$ and $\xi_a^i$. We compare numerical
solutions for the swarm center of mass radius and angular velocity with simulations of a swarm
with Erd{\"o}s-Renyi communication network structure (see Fig.~\ref{fig:rotstate}). Solution curves
for $\xi_{\rho}^i$ and $\Delta \theta_i$ for different
values of $\bar{a}$ are shown in Fig.~\ref{fig:twopopthm}; a comparison with simulation results
is shown in Fig.~\ref{fig:twopopthm2}. 
The slight discrepancy in the rotating state radius in
Fig.~\ref{fig:rotstate} and in $\xi_\rho^i$ in Fig.~\ref{fig:twopopthm2} is understood as follows. 
The radius of the center of mass computed from (\ref{eq:wsqr}) and (\ref{eq:wrho}) assumes
that agent positions' deviate only slightly from the center of mass. However, as the mean coupling 
coefficient decreases the agents become spread out over an extended arc  (as seen in 
Fig.~\ref{fig:rotex}) and our assumption becomes invalid. In this `arc'  configuration the center 
of mass of the swarm moves closer towards the center of rotation than theory predicts.
The analogue to a system with perturbed coupling coefficient
breaks down here; for a globally-connected
swarm with decreasing coupling coefficient $a$, the rotating state disappears
when the system crosses the curve $\bar{a} \tau^2 = 2$, where the rotating
state radius diverges. The swarm then transitions to a translating state. 
It is, however, remarkable, that the mean-field
analysis captures so much of the overall swarm behavior even as the coupling
degree is significantly decreased.

\section{Conclusion}

In this paper we have analyzed the collective motion patterns of a swarm with Erd{\"o}s-Renyi
communication network structure, using a mean-field approach from statistical physics, with the
assumption that the number of agents goes to infinity. We derived bifurcation diagrams demarcating
regions of different
collective motions, for different values of mean degree in the communication network. We showed that
behaviors described in \cite{Romero2012} for the globally-coupled swarm, namely translation, ring
state, and rotation, persist as communication links are broken, even though the bifurcation curves are
shifted as coupling degree of the network decreases.

We derive expressions for the speed of the swarm in the translating state as a function of time
delay and coupling coefficient; for the mean radius and angular velocity of agents in the ring state;
and for the angular velocity, and individual radii and phase offsets for individual agents in the
rotating state. We have verified these calculations with simulations of the full-swarm dynamics. It
is remarkable that our model reduction, which starts with $N$ second-order delay-differential
equations and yields one equation of the same type, is able to quantitatively capture so many
aspects of the full swarm dynamics, even as the coupling degree of agents within the swarm is
significantly decreased.

In the case that many agents are coordinating together, limited communication bandwidth makes 
all-to-all communication infeasible, and may lead to significant communication delays. By dropping 
the requirement for all-to-all communication used in our previous work, the current paper brings us 
one step closer to a possible implementation of swarming control algorithms for very large 
aggregates of agents. Understanding the natural emerging dynamics of the system in these 
circumstances allows us to exploit them when designing controls for swarming applications.

In future work, we will study the effects of using nearest-neighbor communication rather than the random
Erd{\"o}s-Renyi communication network used here. We will verify our results in the lab using a swarm of 
quadcopters operating in a mixed real-virtual environment, in which arbitrarily large numbers of virtual 
agents can be simulated to interact with the group of real vehicles. We will test how our results scale 
with the number of agents in the network, and apply parametric control for dynamic pattern-switching.

\section*{Acknowledgments}

This research was performed while KS held a National Research Council Research Associateship Award at the U.S. Naval Research Laboratory. This research
is funded by the Office of Naval Research contract no. N0001412WX2003 and the
Naval Research Laboratory 6.1 program contract no. N0001412WX30002. LMR is a
post-doctoral fellow at Johns Hopkins University supported by the National
Institutes of Health.

\bibliographystyle{ieeetr}

\end{document}